\newcommand{\optbar}[1]{\shortstack{{\tiny (\rule[.4ex]{1em}{.1mm})}
  \\ [-.7ex] $#1$}}

\documentstyle[sprocl,epsfig]{article}

\def\be{\begin{equation}}
\def\ee{\end{equation}}
\def\bea{\begin{eqnarray}}
\def\eea{\end{eqnarray}}

\begin{document}

\begin{titlepage}

\begin{flushright}
HUPD-9923\\
hep-ph/0001321
\end{flushright}

\vspace{1.5cm}

\boldmath
\begin{center}
\Large\bf CALCULATION OF DIRECT CP VIOLATION IN B DECAYS

\end{center}
\unboldmath

\vspace{1.2cm}

\begin{center}
 Cai-Dian L\"u\footnote{JSPS research  fellow.}

{\sl  

Department of Physics, Hiroshima University, 
Higashi-Hiroshima 739-8526, Japan
\\E-mail: lucd@theo.phys.sci.hiroshima-u.ac.jp

}
\end{center}

\vspace{1.3cm}

\begin{center}
{\bf Abstract}\\[0.3cm]
\parbox{11cm}{
Using the 
generalized factorization approach, we calculate
the CP asymmetries of charmless B decays. 
A number of decay channels has
large CP asymmetries, which can be measured in the B factories.

}

\end{center}

\vfill

\begin{center}
{\sl Invited talk given at the\\
3rd International Conference on B Physics and CP Violation (BCONF99)\\
Taipei, Taiwan, 3--7 December 1999\\
To appear in the Proceedings}
\end{center}

\vspace{1.5cm}

\end{titlepage}

\newpage

\setcounter{page}{1}


\title{CALCULATION OF DIRECT CP VIOLATION IN B DECAYS}

\author{ Cai-Dian L\"u}

\address{Department of Physics, Hiroshima University, 
Higashi-Hiroshima 739-8526, Japan
\\E-mail: lucd@theo.phys.sci.hiroshima-u.ac.jp}


\maketitle\abstracts{ Using the 
generalized factorization approach, we calculate
the CP asymmetries of charmless B decays. 
A number of decay channels has
large CP asymmetries, which can be measured in the B factories.}

\section{Introduction}

The sourse of CP violation is one of the unsolved problem in the
standard model (SM).
The richness of charmless hadronic decays of B meson provides a good place
for study of CP violation \cite{akl2}. 
When  $B^0$ and $\bar B^0$ decay to  a common 
final state $f$, 
 $B^0$-$\bar B^0$ mixing plays a crucial role.
 It
interferences with the direct CP asymmetries.
 For other decays, 
$B^0$ and  $\bar{B}^0$ decay to different final states,
 for example  ${B}^0 \to K^+\pi^-$, $\bar{B}^0 \to K^-\pi^+$.
No mixing is involved here. They are 
 similar to  charged $B^\pm$ decays. 
CP asymmetry has  no time dependence.
The  direct CP asymmetry is important even if for neutral B meson decays.

If there is only one amplitude contributing to the decay,
 both the strong phase and weak phase
can be factored out.
we have  $\Gamma =\overline \Gamma$. So there is 
 no direct CP asymmetry.
That is the case for D meson decays and B meson going to heavy meson
decays, like $B\to J/\psi K_S$.
If there are two amplitudes, the decay rate of $
\Gamma$ and $\overline \Gamma$ may be different. If the strong phase
difference between the two amplitudes $M_1$ and $ M_2$ is
 not zero ($\delta_{12} \not = 0$) and the weak
phase difference of the two amplitudes is also non zero
  ($\phi_{12} \not =0$),
we have  $\Gamma \not =\overline \Gamma$.
The direct CP asymmetry is 
 \begin{eqnarray}
A_{CP}=\frac{2r\sin \delta_{12} \sin \phi_{12}}{1+r^2+2r\cos \delta_{12}
\cos \phi_{12}},
\end{eqnarray}
where $r=|M_2|/|{M_1}|$.
$A_{CP}$ depends on  $2r/(1+r^2)$, 
 $\sin \delta_{12}$ and  $\sin\phi_{12}$.
 If one of the three parameters is small, then $A_{CP}$ is small.
In many decays, we do not have all these conditions, then there
is no sizable direct CP violation. However, most  charmless decays
have large values for  $2r/(1+r^2)$, where 
 $M_1$ is  tree amplitude and  $M_2$ is penguin amplitude.
  Furthermore, the CKM parameters for the tree diagram
and penguin diagrams are different providing weak phase differences. 

 Direct CP asymmetries
 require an interference between two 
amplitudes involving both a  CKM phase
  and a final state  strong
interaction phase difference. 
The weak phase difference arises from the superposition of 
  penguin contributions and the 
 tree diagrams.
The strong-phase difference
arises through 
the  perturbative penguin diagrams (hard
final state interaction),
or  non-perturbatively 
(soft final state interaction).
The soft part is not important which is
shown in some model calculations \cite{fsi}.
There are also some
other contributions, such as annihilation diagram and Soft final
state interaction.
Mostly, their contributions to branching ratios are small \cite{akl1}. 
Probably   their contribution to  $A_{CP}$ is  also  small.
This is also shown in some model calculations \cite{fsi}.

The method of  Isospin or SU(3) symmetry \cite{gronau} which
requires  a set of measurements
 to solve the uncertainties is sometimes difficult for experiments.
We estimate these strong phases in  specific models, such like
the generalized factorization approach, which
can be tested by experiments.

\section{  CP Violation Classification and Formulae}

For charged $B^\pm$ decays the CP-violating asymmetries are
defined as \cite{akl2}
 \begin{equation}
  A_{CP} = \frac{  \Gamma ( B^+ \to  f^+)-\Gamma ( B^- \to  f^-) }{
\Gamma ( B^+ \to  f^+)+  \Gamma ( B^- \to  f^-) }.
\end{equation}
The charged modes are self-tagged decay channels for experiments.
They are easy to be measured.
For  $B^0$ decays, more complication is from the $B^0-\overline B^0$ mixing.
The CP-asymmetries may be  time-dependent, if the final states are the 
same for $B^0$ and $\overline{B^0}$
 \begin{eqnarray}
  A_{CP}(t)&=& \frac{ \Gamma (B^0(t) \to f) -\Gamma (\overline  B^0(t) \to  
 f)}{
 \Gamma (B^0(t) \to f) +\Gamma ( \overline  B^0(t) \to   f)}\\\\
&\simeq& a_{\epsilon'} \cos (\Delta m t) + a_{\epsilon +\epsilon '} 
\sin (\Delta m t).
\end{eqnarray}
Here the direct CP violation parameter $a_{\epsilon'}$  is defined as
  \begin{eqnarray}
a_{\epsilon '} &=&  A_{CP}^{dir}=
 \frac{  \Gamma ( B^0 \to  f)-\Gamma ( \bar B^0 \to   f) }{
\Gamma ( B^0 \to  f)+  \Gamma (\bar B^0 \to f) },
\end{eqnarray}
which is the same defination as the charged B decays.
And
 $a_{\epsilon +\epsilon '} $ is mixing-induced CP violation \cite{akl2}.
In this note we will concentrate on the direct CP asymmetries.

   \subsection{  $b \to s$ ($\bar{b} \to 
\bar{s}$), transitions}

First we parametrize the decay amplitude like this way \cite{akl2}
\begin{eqnarray}
{\cal M}&=&  T \xi _u -P_t \xi_t  e^{i\delta_t} -P_c \xi_c  e^{i\delta_c}
-P_u \xi_u  e^{i\delta_u},\nonumber\\
\overline{\cal M} &=&  T \xi^* _u -P_t \xi_t^*  e^{i\delta_t} -P_c \xi_c^*  
e^{i\delta_c} -P_u \xi_u^*  e^{i\delta_u},\label{mm}
\end{eqnarray}
where  $\xi_i=V_{ib}V^*_{is}$. $T$ and $P_i$ are the tree and $i$ 
($i=u,c,t$)
quark penguin contributions, respectively.
Working in SM, 
we can use the unitarity relation  $\xi_c=-\xi_u-\xi_t $
to simplify the above equation,
 \begin{eqnarray}
{\cal M}&=&  T \xi _u -P_{tc} \xi_t  e^{i\delta_{tc}} 
-P_{uc} \xi_u  e^{i\delta_{uc}},\nonumber\\
\overline{\cal M} &=&  T \xi^* _u -P_{tc} \xi_t^*  e^{i\delta_{tc}}   
 -P_{uc} \xi_u^*  e^{i\delta_{uc}},\label{msim}
\end{eqnarray}
where we define
\begin{eqnarray}
P_{tc}  e^{i\delta_{tc}}&=&  P_t e^{i\delta_t} -P_c e^{i\delta_c} ,\nonumber\\
P_{uc}  e^{i\delta_{uc}}&=&  P_u   e^{i\delta_u}-P_c e^{i\delta_c}.
\end{eqnarray}
 Thus, the direct CP-violating asymmetry is
  \begin{equation}
A_{CP} \equiv a_{\epsilon^\prime}=
\left( |\overline{\cal M} |^2-|{\cal M}|^2 \right)/
\left( |{\cal M} |^2+ |\overline{\cal M}|^2 \right)={A^-}/{A^+}~,
\end{equation}
where
\begin{eqnarray}
A^- &=& 2 TP_{tc} |\xi _u^* \xi_t| \sin \phi_3 \sin \delta_{tc}
+ 2 P_{tc} P_{uc} |\xi _u^* \xi_t| \sin \phi_3
 \sin (\delta_{uc}-\delta_{tc}),
\label{aminus}
\end{eqnarray}
\begin{eqnarray}
A^+ &=& (T^2+P_{uc}^2) |\xi_u|^2 + P_{tc}^2  |\xi_t|^2 
- 2 P_{tc} P_{uc} |\xi _u^* \xi_t| \cos \phi_3
 \cos (\delta_{uc}-\delta_{tc})
\nonumber\\
&&- 2 TP_{uc} |\xi _u |^2\cos \delta_{uc}
+2 TP_{tc} |\xi _u^* \xi_t| \cos \phi_3 \cos \delta_{tc}.
\label{aplus}
\end{eqnarray}

\begin{figure}
  \begin{center}
    \epsfig{file=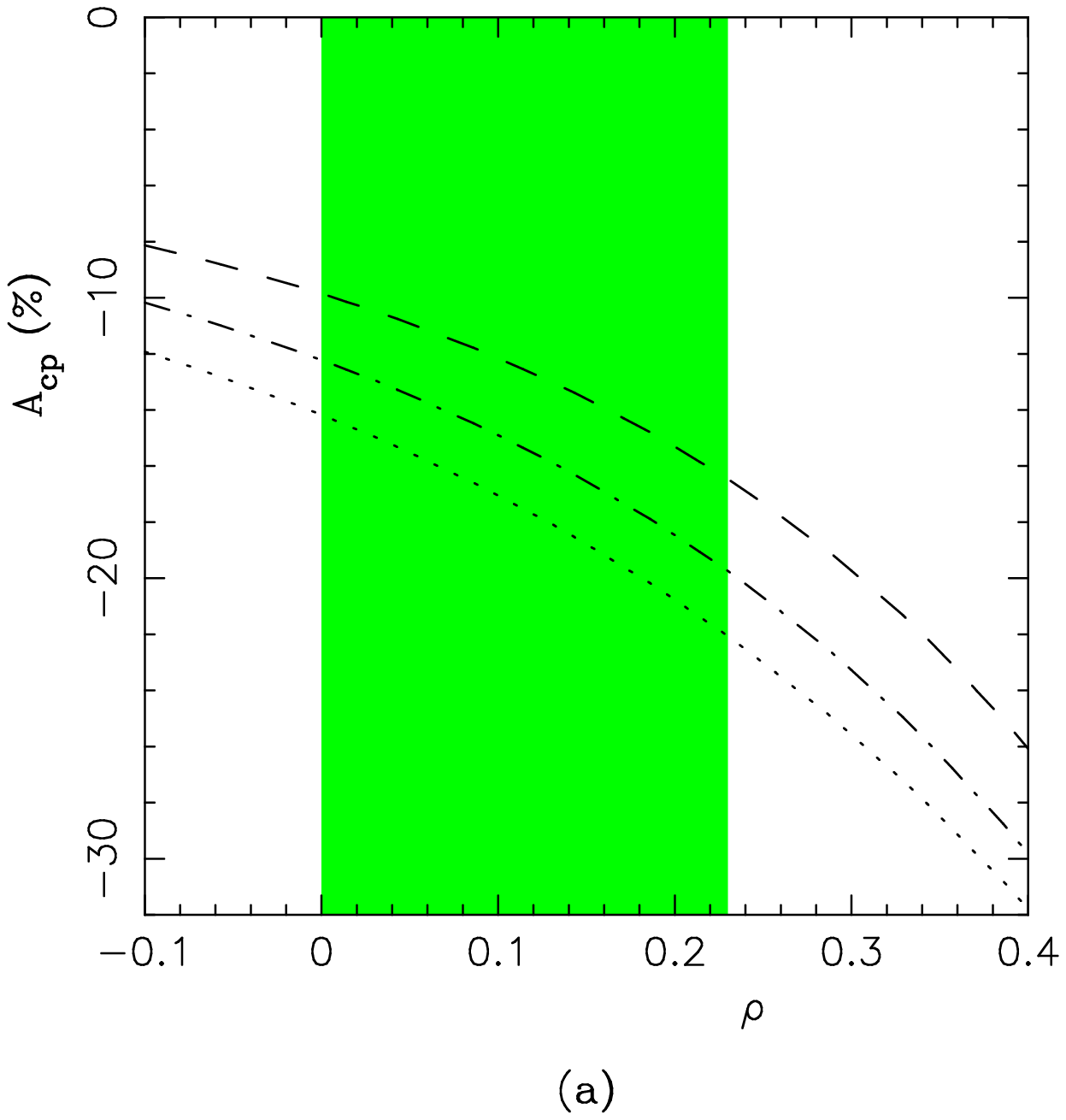,bbllx=5cm,bblly=7cm,bburx=18cm,bbury=19.3cm,%
width=5cm,height=4.4cm,angle=0}
    \epsfig{file=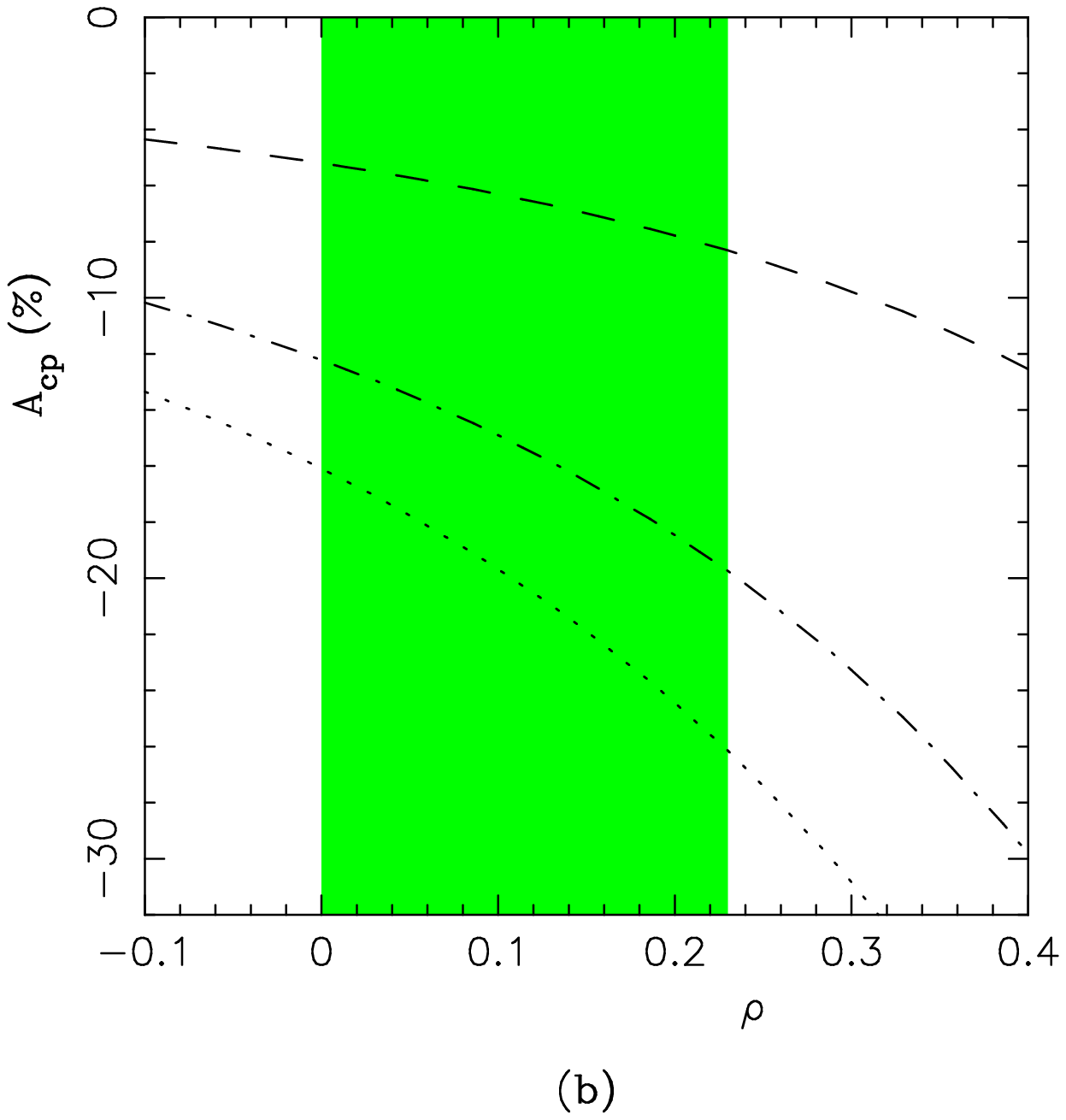,bbllx=5cm,bblly=7cm,bburx=18cm,bbury=19.3cm,%
width=5cm,height=4.4cm,angle=0}
\caption{CP-violating Asymmetry $A_{CP}$ in $ {B^0}\to
K^{*+}\pi^-$
decay  as a function of the CKM parameter $\rho$.
(a) $k^2=m_b^2/2$. The dotted, dashed-dotted and dashed curves correspond to
the CKM parameter values $\eta=0.42$, $\eta=0.34$ and $\eta=0.26$,
respectively.
(b)  $\eta=0.34$. 
The dotted, dashed-dotted and dashed curves correspond to
$k^2=m_b^2/2+2$  GeV$^2$, $k^2=m_b^2/2$ and $k^2=m_b^2/2-2$ GeV$^2$, 
respectively.} \label{cc2}
  \end{center}
\end{figure}

First, we note that   $|\xi_u| \ll | \xi_t|\simeq
|\xi_c|$, with an upper bound  $|\xi_u|/| \xi_t| \leq 0.025$.
In some channels, such as 
$B^+ \to K^+ \pi^0$, $K^{*+} \pi^0$,
$K^{*+} \rho^0$, $B^0 \to K^+ \pi^-$, $K^{*+} \pi^-$, $K^{*+} \rho^-$,
  $|P_{tc}/T|$ is of $O(0.1)$, 
 $|P_{uc}/P_{tc}| =O(0.3)$.
The CP-violating asymmetry in this case is 
 \begin{equation}
  A_{CP} \simeq \frac{2z_{12} \sin \delta_{tc}  \sin \phi_3}
{1+ 2 z_{12} \cos \delta_{tc} \cos \phi_3 +z_{12}^2},\label{app1}
\end{equation}
where $z_{12}=|\xi_u/\xi_t|\times T/P_{tc}$.
We show the CP asymmetry of $B\to K^{*+} \pi^-$ as an example in Figure 1.
It is easy to see that, there may be large CP asymmetries in this decay 
channel. Besides the CKM parameter $\rho$ and $\eta$, 
the CP asymmetry is also sensitive to the gluon momentum $k^2$, which 
is related to the size of strong phase. If $k^2$ is known, the strong phase
is predictable, we may use $A_{CP}$ to determine $\sin\phi_3$.
The first 6 channels of Table 1 are this kind of decays.
Two of them are reported from CLEO Collaboration with large error-bars 
\cite{smith}.
The central values are far away from the theoretical predictions.
If more data suport this, we may expect new physics signals here.

\begin{table}[ht]
\caption{CP-rate asymmetries $A_{CP}$ and 
branching ratios for some
 $B  \to h_1 h_2$ decays, 
updated for the central values
of the CKM fits $\rho=0.20$, $\eta=0.37$   
and the factorization model parameters $\xi=0.5$ and $k^2=m_b^2/2\pm 2$
GeV$^2$. }
\begin{center}
\begin{tabular}{|c|c|c|l|}
\hline
Decay Modes & \raisebox{0pt}[13pt][7pt]{$A_{CP}$-Exp.(\%)}&
\raisebox{0pt}[13pt][7pt]{$A_{CP}
(\%)$}&\raisebox{0pt}[13pt][7pt]{$BR(\times 10^{-6})$}\\
\hline
$B^\pm \to K^\pm \pi^0$ &$ 29\pm 23$    &$ -7.7^{-2.2}_{+4.0}$ & $10.0$\\
\hline
$B^\pm \to K^{*\pm} \pi^0$ &$ -$ &$-14.4^{-4.4}_{+8.2}$ &$4.3$\\
\hline
$B^\pm \to K^{*\pm} \rho^0$ &$ -$ &$ -13.5^{-4.0}_{+7.5}$ &$4.8$\\
\hline
${B^0} \to K^+ \pi^-$ & $4\pm 16$   &$ -8.2^{-2.3}_{+4.3}$ &$14.0$\\
\hline
${B^0} \to  K^{*+}\pi^-$ & $ -$ & $ -17.2^{-5.5}_{+9.8}$
& $ 6.0$\\
\hline
${B^0} \to K^{*+} \rho^-$ &$ -$ & $ -17.2^{-5.5}_{+9.8}$
&$5.4$\\
\hline
$B^\pm \to K^0_S \pi^\pm$ &$ -18\pm 24$    &$ -1.4^{-0.1}_{+0.1}$ & $14.0$\\
\hline
$B^\pm \to \eta \pi^\pm $ &$ -$   &$ 9.3^{+1.9}_{-4.1}$ & $5.5$ \\  
\hline
$B^\pm \to \eta^\prime \pi^\pm $&$ -$   &$ 9.4^{+2.1}_{-4.5}$ & $3.7$\\
\hline
$B^\pm \to \eta \rho^\pm $ &$ -$ &$3.1^{+0.7}_{-1.7}$ &$8.6$\\
\hline
$B^\pm \to \eta^\prime \rho^\pm $ &$ -$  &$3.1^{+0.7}_{-1.8}$ &$6.2$\\
\hline
$B^\pm \to \rho^\pm \omega$ & $ -$ & $7.0^{+1.5}_{-3.4}$ & $ 21.0$\\
\hline
$B^\pm \to \eta^\prime K^\pm$& $ -3\pm 12$ &$-4.9^{-1.2}_{+2.1} $ & $23.0$\\
\hline
$B^\pm \to \pi^\pm \omega$ &$ 34\pm 25$ &$7.7^{+1.7}_{-3.7}$ &$9.5$\\
\hline
   \end{tabular} \end{center} 
\end{table}  

  There are some decays with  vanishing
 tree contributions ($T=0$), such as 
 $B^+ \to \pi^+ K_S^0$, $\pi^+ K^{*0}$, $\rho^+ K^{*0}$. 
 Then for these decays,
the CP-violating asymmetry is 
 \begin{equation}
  \label{cps2}
  A_{CP} \simeq 2\frac{P_{uc}}{P_{tc}}\left |\frac{\xi_u}{\xi_t}\right |
 \sin (\delta_{uc}-\delta_{tc})  \sin \phi_3.
\end{equation}
Without the tree  contribution, the suppression due to both 
 $P_{uc}/P_{tc}$ and
 $|\xi_u/\xi_t|$ is much stronger.
The CP-violating asymmetries are only
around 
 $-$($1$-$2$)\%. 

Some estimates of the channel $B^+\to \pi K_S^0$ show that
even including the annihilation and soft final state interaction, 
the CP asymmetry of this decay is still small \cite{fsi}.
 This means that this channel is clean
for new physics to show up.
In table 1, we can see that CLEO's central value of this decay indicates
a large CP asymmetry maybe possible.

  \subsection{ $b\to d$ ($\bar b \to \bar d$) transitions}

Similarly to the $ b\to s$ transition, we can define the CP asymmetry as
 \begin{equation}
A_{CP} =A^- /A^+,
\end{equation}
where
\begin{eqnarray}
A^- &=& -2 TP_{tc} |\zeta _u^* \zeta_t| \sin \phi_2 
\sin \delta_{tc}
- 2 P_{tc} P_{uc} |\zeta _u^* \zeta_t| \sin \phi_2
 \sin (\delta_{uc}
-\delta_{tc}),
\label{aminusd}
\end{eqnarray}
\begin{eqnarray}
A^+ &=& (T^2+P_{uc}^2) |\zeta_u|^2 + P_{tc}^2  |\zeta_t|^2 
- 2 P_{tc} P_{uc} |\zeta _u^* \zeta_t| \cos \phi_2 \cos 
(\delta_{uc}-\delta_{tc})
\nonumber\\
&&- 2 TP_{uc} |\zeta _u |^2\cos \delta_{uc}
+2 TP_{tc} |\zeta _u^* \zeta_t| \cos \phi_2 \cos \delta_{tc},
\end{eqnarray}
with  $\zeta_i=V_{ib}V^*_{id}$,
 and again we have used CKM unitarity relation
 $\zeta_c = -\zeta_t-\zeta_u $.

For the tree-dominated decays, such as 
$B^+\to \pi^+ \eta ^{(\prime)}$, $\rho^+ \eta ^{(\prime)}$, $\rho^+ \omega$,
the relation $P_{uc} < P_{tc} \ll T$ holds. The CP asymmetry is
 \begin{equation}
 A_{CP} \simeq \frac{-2z_{1} \sin \delta_{tc}  \sin \phi_2}
{1+ 2 z_{1} \cos \delta_{tc} \cos \phi_2},\label{app3}
\end{equation}
with $z_{1}=|\zeta_t/\zeta_u|\times TP_{tc}/T'^2$,
and $T'^2\equiv T^2  - 2 TP_{uc}\cos \delta_{uc}$.
The CP asymmetries  are proportional to $\sin \phi_2$. They are large
enough for the experiments to detect them. The theoretical predictions
of these decays are shown in table 1.

For the
decays with a vanishing tree contribution ($T=0$), such as  
$B^+ \to K^+  K_S^0$, $K^+ \bar K^{*0}$, $K^{*+} \bar K^{*0}$, 
the CP-violating asymmetry is approximately  proportional to  $\sin\phi_2$
again,
 \begin{equation}
 A_{CP} = \frac{-2z_{3} \sin (\delta_{uc}-\delta_{tc})  \sin 
\phi_2}
{1- 2 z_{3} \cos (\delta_{uc}-\delta_{tc})  \cos \phi_2
+z_3^2},\label{app4}
\end{equation}
with  $z_3=|\zeta_u/\zeta_t|\times P_{uc}/P_{tc}$.
As the  suppressions from  
$|\zeta_u/\zeta_t|$ and $|P_{uc}/P_{tc}|$  
are not very big, the CP-violating asymmetry can again be
of order  $(10$-$20)\%$. 
Unfortunately, these channels have  smaller branching ratios
\cite{akl2,akl1}.

More charmless decay channels are discussed in ref.\cite{akl2}.
 Some of them are more complicated than the
ones we discussed above. There are also 
   some 
other  interesting decays like
$B\to K^* \gamma$,
$B \to D \pi \ell \nu$,
$B\to \pi\pi \ell \nu$ \cite{kim}, etc.
They have small CP asymmetries in SM.
 They are sensitive to    new physics.

\section{Models of Calculation}

In the  Factorization Approach\cite{akl1,cheng}, the two body $B$ meson decays
can be factorized as two products:
$$ C_i ~\langle P_1 P_2 | O_i | B \rangle 
= C_i ~\langle P_1 |J_\mu | 0 \rangle 
~~\langle  P_2 | J^\mu | B \rangle ,$$
 where $C_i$ is the corresponding Wilson coefficients. 
The second factor on the right side of the equation is proportional to
the meson decay constant. The last term is the corresponding form factors.

The strong-phase differences 
arise through  Bander-Silverman-Soni Mechanism (BSS) \cite{bss}.
In this picture,
the  perturbative penguin diagrams involving 
	 charm and up quark loops, where the light quarks can be on 
mass shell, providing the strong phases. They are
  mostly sensitive to the gluon momentum $k^2$.
For numerical calculations, we use  $k^2 =m_b^2/2 \pm 2 GeV ^2$.

In the 
  perturbative QCD approach (pQCD) \cite{li},
we need one hard gluon  connecting the spectator quark.
Strong phases are from the  non-factorizable diagram and
annihilation diagram, where the
innner quark or gluon propagator can be on mass shell.
The pQCD approach is based on  factorization, 
and goes one step further. In pQCD, we can calculate
 annihilation diagrams and also the
non-factorizable contributions. The
	  $k^2$ of gluon is well defined in this approach.
We have calculated the $B\to \pi \pi$ decays in this approach, and 
the results compared with the factorization approach in Table 2.

\begin{table}\caption{Direct CP asymmetries of
$B\to \pi\pi$ decays in factorization approach
and perturbative QCD approach (Preliminary) \protect\cite{luy}.}\begin{center}
\begin{tabular}{l|rr}
Channel & $A^{direct}_{CP}$  (Factorization) & $A^{direct}_{CP}$ 
(pQCD) \\
\hline
$B\to \pi^+\pi^-$ &  $6.9\%^{+2}_{-4}$ 
		  &  -1.4\%\\
\hline
$B^+\to \pi^+\pi^0$ & $0.1\%^{+0.1}_{-0.1}$ &
		   0.02\%\\
\hline
$B^0\to \pi^0\pi^0$  & $-15\%^{-7}_{+14}$ &
		    -60\%\\\hline
\end{tabular}\end{center}
\end{table}

 \section{  Summary}

 The recently measured direct CP asymmetries  for
$B^0 \to K^+\pi^-$, $B^+ \to K^+ 
\eta^\prime$, $B^0 \to K^+ \pi^-$, $B^+ \to \pi^+ K^0$, and $B^+ \to 
\omega \pi^+$ are encouraging news for direct CP violation in B decays,
although the signal is not significantly excess background.

 CP-asymmetries of
$A_{CP}(K^\pm\eta^\prime)$, $A_{CP}(\pi^\pm K_S^0)$ and $A_{CP}(\rho^\pm 
\optbar{K^{*0}})$
  are small,
but stable against variation in $N_c$, $k^2$ and $\mu$. 
 CP-asymmetries well  over $10\%$ in these decay modes
will be a sign of  new physics.
The decay channels of $B\to 
K^{*\pm} \pi^\mp$, $K^{*\pm} \pi^0$,
$K^{*\pm} \eta$, 
$K^{*\pm} \eta^\prime$, $K^{*\pm} 
\rho^\mp$ and $K^{*\pm} \rho^0$,
have measurably  large
CP-violating asymmetries.
A good measurement of
the CP-asymmetry in any one of these decays could be used to 
determine  $k^2$.
Such that the theoretical predictions of all other channels
make sense. 
We also hope that the perturbative QCD approach could solve 
the remaining uncertainties in the factorization approach.
With the two B factories and other hadronic machines, 
 a number of decays is  going to be measured soon.

\section*{Acknowledgments}
The author thanks A. Ali and G. Kramer for collaboration on the main topic
discussed in this report. We thank the organizer H.Y. Cheng and 
W.S. Hou for a fruitful BCP3 conference. We acknowledge the
Grant-in-Aid for Scientific Research
on Priority Areas (Physics of CP violation) and JSPS for support.

\end{document}